\documentstyle[epsf,twocolumn,aps]{revtex}
\newcommand{\eq}{\begin{equation}}
\newcommand{\ee}{\end{equation}}
\newcommand{\eqa}{\begin{eqnarray}}
\newcommand{\eea}{\end{eqnarray}}

\def\dg{\delta g}
\def\de{\Delta E}
\def\ef{E_F}
\def\ag{\langle g\rangle}

\newcommand{\bt}{{\bf T}} 
\newcommand{\btdg}{{{\bf T}^\dagger}} 
 
\newcommand{\btt}{{\bf t}}

\newcommand{\tr}{{\rm Tr}} 
\newcommand{\pprl}{Phys. Rev. Lett. \ } 
\newcommand{\pprb}{Phys. Rev. {B}} 
\begin{document}
\twocolumn[
\hsize\textwidth\columnwidth\hsize\csname@twocolumnfalse\endcsname
\draft
\title{Conductance Correlations Near Integer Quantum Hall Transitions}
%\title{Energy Correlation Length in the Quantum Hall Plateau Transitions}
\author{Bo\v zidar Jovanovi\'c$^{a,b}$ and Ziqiang Wang$^a$}
\address{$^a$ Department of Physics, Boston College, Chestnut Hill, MA 02167}
\address{$^b$ Department of Physics, Boston University, Boston, MA 02215}

\date{\today}
\maketitle

\begin{abstract}
In a disordered mesoscopic system,  the typical spacing between the peaks 
and the valleys of the conductance as a function of Fermi energy $E_F$ 
is called the conductance energy correlation range $E_c$. Under the 
ergodic hypothesis, the latter is determined by the half-width of the 
ensemble averaged conductance correlation function:  
$F= \langle \delta g(E_F) \delta g(E_F + \Delta E) \rangle$. 
In ordinary diffusive metals, $E_c\sim D/L^2$, where $D$ is the diffusion 
constant and $L$ is the linear dimension of the phase-coherent sample.  
However, near a quantum phase transition driven by the location 
of the Fermi energy $E_F$, the above picture breaks down. 
As an example of the latter, we study, for the first time, the conductance 
correlations near the integer quantum Hall transitions  
of which $E_F$ is a critical coupling constant. We point out that
the behavior of $F$ is determined by the interplay between the
static and the dynamic properties of the critical phenomena.
\end{abstract}
]
%\narrowtext

The quantum interference effects in disordered phase-coherent systems
belong to the mesoscopic physics \cite{ucfreview,ucf}. 
A phase-coherent sample is
one in which the phase-coherence length $L_\phi$ is larger
than the sample size $L$. Thus mesoscopic physics naturally appears
in small systems of mesoscopic dimensions, usually in nanostructures.
Mesoscopic physics is also important in 
systems large enough to exhibit macroscopic quantum phase transitions.
The reason is that $L_\phi$, being the cutoff for the critical fluctuations, 
diverges as the temperature $T$ approaches zero at such transitions.
An important example of the latter is a two dimensional electron gas (2DEG) 
close to the transition between two quantized Hall plateaus.
The mesoscopic fluctuations of the conductance in this case has
been studied recently both experimentally and theoretically
\cite{cobden,wjl,chofisher,shanhui,bhatt,ybkim,jw}.

The essential physics in the mesoscopic regime is the lack of self-averaging
in the transport properties. Sample specific, reproducible fluctuations
in the conductance become observable at low temperatures.
From a theoretical point of view, the basic statistical properties
of conductance fluctuations are determined by the conductance
correlation function \cite{ucf},
\eq
F(E_F,\de)=<\dg(E_F+\de) \dg(E_F)>,
\label{f}
\ee
where $\dg$ is the deviation from the impurity averaged conductance,
{\it i.e.} $\dg(E_F)=g(E_F)-<g(E_F)>$. 
Note that the conductances in $F$ are to be evaluated at {\it two different
Fermi energies} separated by the amount $\de$.
In general, $F$ can have another argument $\Delta B$, representing
the magnetic field correlation of the conductance, which
we shall not consider here.

For $\de=0$, Eq.~(\ref{f}) gives the variance of the conductance,
${\rm Var}(g)=F(0)=<\dg^2>$. In the rest of the
paper, we measure the conductance in units of
$e^2/h$ and the variance in units of $(e^2/h)^2$.
Under the ergodic hypothesis, the sample to sample fluctuations
are analogous to the fluctuations of the conductance as a
function of the Fermi energy. In this case, the typical spacing
of the peaks and valleys in the conductance as a function of energy
in a specific sample, usually called the energy correlation range $E_c$,
is determined by the half-width of the ensemble averaged 
conductance correlation function $F(\de)$ in Eq.~(\ref{f}), 
{\it i.e.} $E_c=\de_{\rm1/2}$.

In ordinary mesoscopic disordered metals in the diffusive regime, 
the disorder-averaged conductance, $\ag$, can vary by orders
of magnitude, but the variance of the conductance
assumes a universal value of order one \cite{ucf}.
Moreover, the energy correlation range of the conductance is
given by $E_c\approx \hbar\pi^2 D/L^2$, where $D$ is the diffusion
constant and $L$ is the linear dimension of the system. It is
important to emphasize that in the diffusive regime, the conductance
correlation in Eq.~(\ref{f}) is only a function of the energy difference
$\de$ and is independent of $E_F$ \cite{ucf}. $E_c$ in this case corresponds
to the inverse diffusion time across the sample in the current direction
which is often referred to as the Thouless energy. This is the characteristic
Fermi energy difference beyond which the paths of two injected electrons 
seize to be phase coherent, giving rise to significant difference
in the conductances.

However, as we shall show in this paper, 
in the critical regime of a quantum phase transition (QPT)
that is {\it driven by the location of the Fermi energy} instead of
correlation strengths, the above picture breaks down. 
The primary reason is that, in this case,
the Fermi energy is a critical coupling constant that
controls the proximity to a quantum critical point. As a result,
the conductance correlation function in Eq.~(\ref{f}) is determined by the 
critical properties associated with the QPT.
We shall focus on the QHE in which the transitions between
the quantized Hall plateaus as a function of the magnetic field
is driven by the location of the Fermi energy of the disordered
2D electron system.

It is well known that the quantum Hall transition (QHT) 
is a continuous zero temperature phase transition at a single extended state
energy $E^*$ between two adjacent Hall plateaus \cite{huckreview}. The critical
singularity at the QHT is described by a single divergent length,
the localization length, $\xi(E_F)\propto\vert E_F-E^*\vert^{-\nu}$,
as $E_F$ approaches $E^*$. Here $\nu$ is
the localization length exponent. We shall focus on
the critical regime, finite systems of linear dimension $L$,
and the zero temperature limit. In this case, the critical fluctuations
are cutoff by $L$. The width of the critical regime shrinks with 
increasing $L$ according to $L^{-1/\nu}$. The
transition thus acquires a finite width $W\propto L^{-1/\nu}$.
The conductance in the transition regime is dominated
by phase coherent transport and thus exhibits mesoscopic phenomena.
What is different from ordinary diffusive metals is the proximity
to the quantum critical point. If the typical spacings between the
peaks and valleys in $g(E_F)$, {\it i.e.} the energy correlation
range $E_c\sim L^{-\alpha}$, a large number of oscillations would
appear within the critical region $W$ so long as $\alpha>1/\nu\simeq0.42$,
which will be shown to be the case below.
In this regime, the ergodic hypothesis, which is expected to fail
in the plateau phases, remains valid. Moreover,
{\it the conductance correlations are determined by the critical properties
associated with the QHT}. 

We now proceed to write down the scaling form for the correlation
function defined in Eq.~(\ref{f}) near the QHT,
\eq
F(E_F,\de)={\cal F}\left[{L\over\xi(E_F+\de)}, {L\over\xi(\ef)},{L\over
L_\omega(\de)}\right].
\label{fscaling}
\ee
Here $L_\omega$ is the length scale introduced by a finite frequency.
The origin of the latter is the following. In calculating the
correlation function of the DC conductances at different energies,
the energy difference $\de$ enters formally as a finite frequency.
This was first pointed out by Lee, Stone, and Fukuyama \cite{ucf} in
their diagrammatic evaluation of $F$ in diffusive metals.
The easiest way to see that $\xi$ must enter the scaling function
is to consider $\de=0$, in which case, Eq.~(\ref{fscaling})
gives the expected result \cite{wjl}: $F(E_F)=\langle\dg^2(E_F)\rangle
={\cal F}[L/\xi(E_F)]$. 
Moreover, both $\xi(E_F+\de)$ and $\xi(E_F)$ must enter as scaling
arguments in ${\cal F}$, because the range of $\de$ wherein
$\xi(E_F+\de)\sim\xi(E_F)$ is given by
$\de\ll {\rm const}\times\xi(E_F)^{-(1+1/\nu)}$, which is very small
in the critical regime and vanishes much faster than the 
transition width $W$.

Eq.~(\ref{fscaling}) shows that in general, in the critical regime
of the QHT, both $E_F$ and $\de$ enter the energy correlation function of
the conductance. More important is the dual-role played by the 
Fermi energy difference. Writing $\xi(E)\sim\vert E-E^*\vert^{-\nu}$
and $L_\omega\sim\vert\de\vert^{-1/z}$ with $z$ the dynamical scaling
exponent, and setting one of the Fermi energy $E_F=E^*\equiv0$,
Eq.~(\ref{fscaling}) becomes,
\eq
F(\de)={\cal F}\left(\vert\de\vert L^{1/\nu},\vert\de\vert L^z\right).
\label{fscaling3}
\ee
Eq.~(\ref{fscaling3}) clearly shows that $\de$ is a coupling constant conjugate
to the static correlation (localization) length, and at the same time,
a quantity analogous to a finite frequency conjugate to the
length scale determined by the dynamical scaling exponent $z$.
As a result, {\it both the static and the dynamic} 
critical properties enter the DC conductance correlation function.
In general one expects that Harris criteria
$\nu z>1$ holds. The two scaling arguments in Eq.~(\ref{fscaling3}) compete and
the correlation function must show a novel crossover from
the regime dominated by static ($\de\sim L^{-1/\nu}$) fluctuations
at large $\de$ to that dominated by dynamic 
($\de\sim L^{-z}$) fluctuations at small $\de$.
Consequently, one expects the energy correlation range $E_c$ 
to interpolate between $E_c\sim L^{-1/\nu}$ at large $\de$ 
and $E_c\sim L^{-z}$ at small $\de$.

We next present a direct numerical calculation of
the conductance correlation function $F$ in Eq.~(\ref{fscaling3})
for an integer QHT in which the effects of electron-electron interactions 
are not considered \cite{lwinter}. In this theoretical
{\it noninteracting analog} of the true integer QHT in real materials,
it is known that $\nu\simeq2.3$ \cite{huckreview} and the
dynamical exponent $z=2$. That $z=2$ 
comes from the energy level spacing in a $d=2$
noninteracting electron system and is consistent with the 
(anomalous) diffusive dynamics known at the noninteracting integer 
QHT \cite{cdnote}.
We will demonstrate that $E_c$
indeed decays as $L^{-1/\nu}$ at large $\de$ and as $L^{-z}$
at small $\de$. Remarkably, the crossover region between the
latter two behaviors is rather broad in $\de$ over which
we find $E_c \sim L^{-1}$.

For convenience, we choose to describe the transport in the integer quantum 
Hall regime using the Chalker-Coddington network model \cite{cc,lwk}.
The latter is a square-lattice of potential saddle points (nodes) where
quantum tunnelings between the edge states of the Hall droplets take
place. With a choice of gauge \cite{cc}, the transfer matrix 
at each node is given by,
\eq 
T_{\rm node}=\pmatrix{\cosh\theta & \sinh\theta\cr 
\sinh\theta & \cosh\theta\cr},
\label{tqh} 
\ee 
with a single real parameter $\theta$.
We have explicitly verified that introducing randomness in $\theta$ does
not change any of our results near the transition in agreement with
the results of Ref.~\cite{lwk}.
Away from the nodes, the edge electrons move along the links (equipotential
contours) with a fixed chirality set by the direction of the magnetic 
field and accumulate random Bohm-Aharonov phases. 
Note that besides the distribution of these random link phases, 
$\theta$ in Eq.~(\ref{tqh}) is the only parameter of the network. 
Changing $\theta$ amounts to
varying the Fermi energy $E_F$ across the QHT.
In the rest of the paper, we will present results in terms of
$E_F$ which is related to the parameter $\theta$ by
the choice of $\theta=\theta_c\exp(E_F-E^*)$ and $\sinh\theta_c=1$
\cite{lwk}.

We have performed large scale numerical calculations
of the two-terminal conductance. To this end,
two semi-infinite ideal leads are attached to the 
left and right ends of the disordered network \cite{wjl,chofisher},
and periodic or open boundary conditions are applied in the transverse 
direction. Let us consider disordered networks having $L$ columns of
nodes and $W/2$ channels. Under such settings, the two-terminal conductance 
of a given sample with a fixed disorder realization
is given by the Landauer formula \cite{fisherlee},
\eq
g(E)={e^2\over h}\tr[\btt^\dagger \btt],
\label{g0}
\ee 
where $\btt$ is the $(W/2)\times (W/2)$ transmission matrix.
In the transfer matrix approach, it is convenient to express $g$
in terms of the ($W\times W$) transfer matrix $\bt$ \cite{pichard},
since the latter is multiplicative across the $L$ columns of scattering 
nodes in the network. Defining the ordered eigenvalues ($\{\lambda_i\}$)
of the symplectic matrix, $H=\btdg\bt$, by $\lambda_i=\exp(2\gamma_i)$
for $i=1,\dots,W$, Eq.~(\ref{g0}) can be written as,
\eq 
g(E)={e^2\over h}\sum_{i=1}^{W/2}{1\over\cosh^2(\gamma_i(E))}. 
\label{g} 
\ee 
Thus the calculation of the two-terminal conductance is transformed into
that of the eigenvalues of the transfer matrix product $H$. It is known
that constructing $\btdg\bt$ by direct matrix multiplications is
numerically unstable when the system size is large. 
We use here the stable numerical algorithm developed recently for 
large scale conductance calculations \cite{wjl}. The details of 
this algorithm have been discussed in Refs.~\cite{jw,pw}.
The basic idea is to maintain the stability of matrix multiplications
using the method of matrix UDR-decomposition, and to extract the 
eigenvalues using the method of orthonormal projection.
Specifically, one can show following a sequence of UDR-decompositions,
the $n$-th power of $H$, can be written as
$
H^n=U_n D_n R_n.
$
In the limit of large $n$, typically less than $15$, 
(i) $U_n$ is a unitary matrix of which 
the columns converge to the eigenvectors of $H$; (ii) $D_n$ 
is a diagonal matrix and the eigenvalues of $H$ is given by 
$D_n D_{n-1}^{-1}$; and (iii) $R_n$ converges to a limiting
right triangular matrix with unity on the diagonals.
We next present the numerical results obtained using this algorithm
on networks with $L=W$, and $L$
up to $128$ in units of the lattice spacing.
\begin{figure}    
%\hspace{1truecm}    
\vspace{-0.5truecm}
\center    
\centerline{\epsfysize=2.6in    
\epsfbox{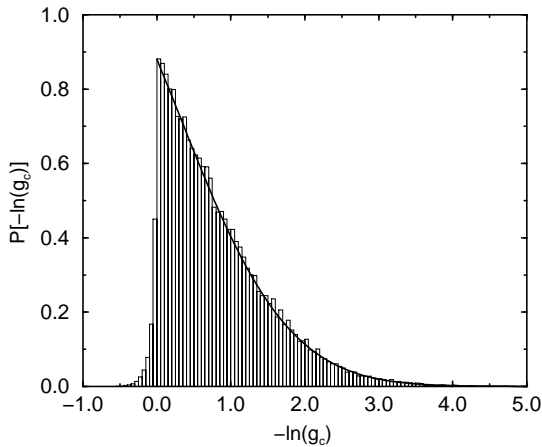}}    
\vspace{-0.5truecm}
\begin{minipage}[t]{8.1cm}    
\caption{ 
Critical conductance distribution showing the skewed log-normal
behavior (solid-line). 
}
\label{fig1}   
\end{minipage}    
\end{figure}
The critical conductance fluctuations at the QHT
have been studied recently\cite{wjl,chofisher}. 
The critical conductance
$g_c$ was found to be broadly distributed between
$0$ and $1e^2/h$ with log-normal characteristics of the central moments. 
Fig.~1 shows the distribution function of $\log(g_c)$
calculated from Eq.~(\ref{g}) for 49,000 disorder realizations at $L=128$.
It shows a remarkable skewed log-normal behavior as a result of
the sharp fall off of $P[g_c]$ close to $1e^2/h$.

In Fig.~2, we plot the conductance as a function of the Fermi energy
in a typical sample for four different sample sizes. 
It is important to note that these reproducible spectra exhibit remarkably 
smooth oscillations with well defined peaks and valleys 
in the transition regime. 
More oscillations appear with increasing system size.
Since the transition width shrinks as $L^{-1/\nu}$,
the typical spacing between the peaks and 
the valleys, {\it i.e.} the correlation range $E_c$, 
must decrease systematically with increasing $L$ faster than as $L^{-1/\nu}$.
These features are very different from the conductance fluctuation spectrum 
in diffusive metals of mesoscopic dimensions.
\begin{figure}    
%\hspace{1truecm}    
\vspace{-0.5truecm}
\center    
\centerline{\epsfysize=2.6in    
\epsfbox{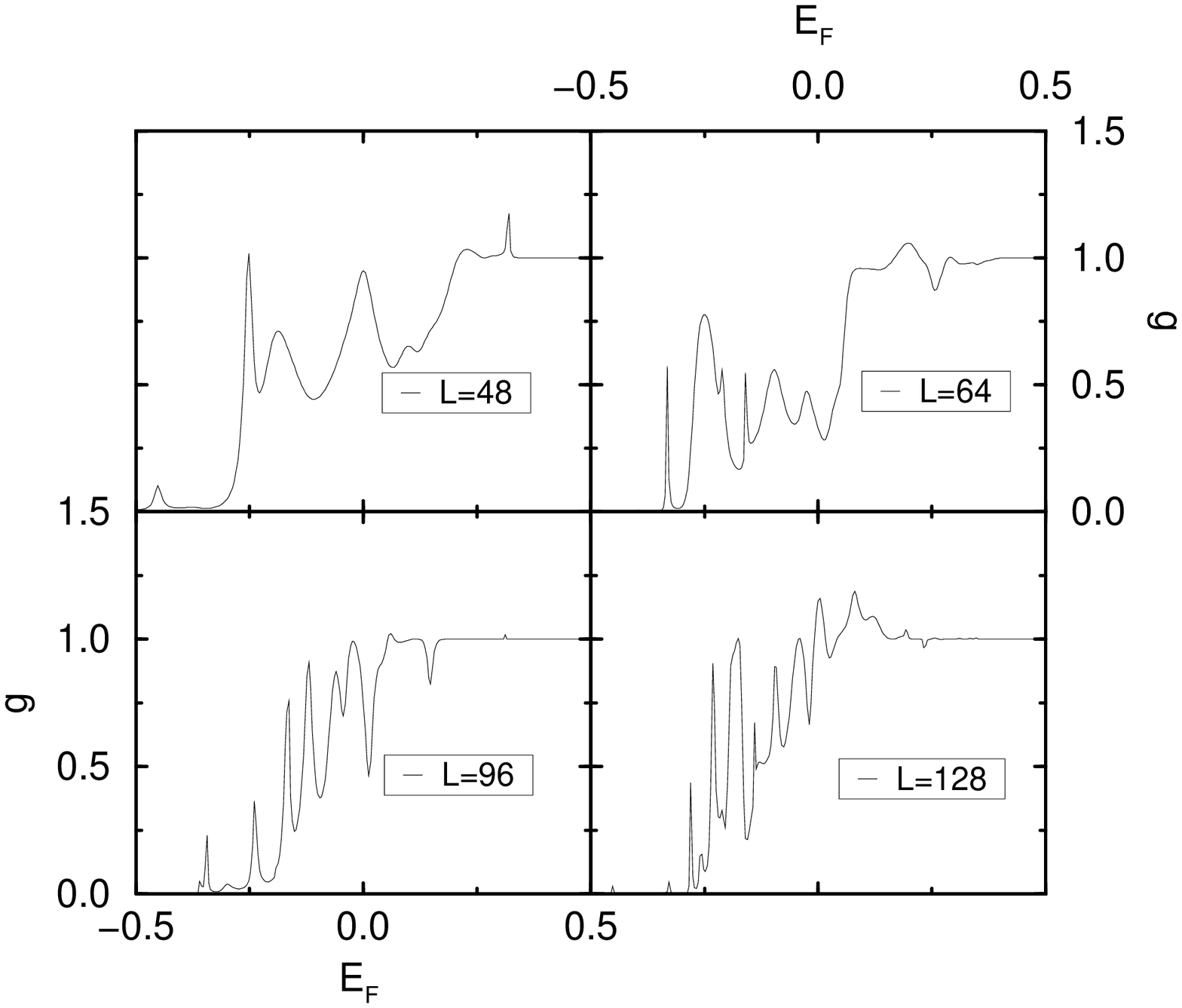}}
\vspace{-0.5truecm}    
\begin{minipage}[t]{8.1cm}    
\caption{ 
Conductance fluctuation spectra in four typical samples of 
$L=48,64,96,128$.
}  
\label{fig2}   
\end{minipage}    
\end{figure}    
From an ensemble of up to 50,000 samples for each $L$, 
having individual fluctuation spectrum exemplified in Fig.~2, 
we calculated the conductance correlation
function, $F(\Delta E)$, defined in Eq.~(\ref{fscaling3}).
In Fig.~3, we plot the normalized conductance correlation function
by the variance of the critical conductance, $F_{\rm n}(\Delta E)\equiv
F(\Delta E)/F(0)$, as a function of
$\Delta E$ for different system sizes.
The half widths $\Delta E_{1/2}$ of each correlation
function curve can be determined to extract the correlation range $E_c$ 
versus $L$. Doing so, we obtained $E_c\sim L^{-1}$.
To understand this rather surprising result, 
we now perform a scaling analysis of the correlation function
in Eq.~(\ref{fscaling3}). 
Notice that the scaling function
has two arguments originated from two different length scales,
$\xi\sim\vert\Delta E\vert^{-\nu}$ and $L_\omega\sim\vert\Delta E\vert^{-1/z}$.
The scaling function should be dominated by the smaller one of the two.
However, both of them are controlled by the same parameter, $\Delta E$, 
{\it i.e.} the difference in the Fermi energy around the critical point.
For the noninteracting QHT, $\nu\simeq2.3$ and
$1/z=0.5$. Thus $L_\omega\ll\xi$ for $\Delta E\ll 1$ whereas $\xi\ll L_\omega$ 
for $\Delta E\gg 1$.
We therefore introduce a new length scale $\lambda(\Delta E)$, which
interpolates between $\xi(\Delta E)$ and $L_\omega(\Delta E)$ as a function
of $\Delta E$, and rewrite Eq.~(\ref{fscaling3}) as
\eq
F(\Delta E)={\cal F}(\lambda(\Delta E)/L).
\label{nfscale}
\ee
\begin{figure}    
%\hspace{1truecm}    
\vspace{-0.5truecm}
\center    
\centerline{\epsfysize=2.6in    
\epsfbox{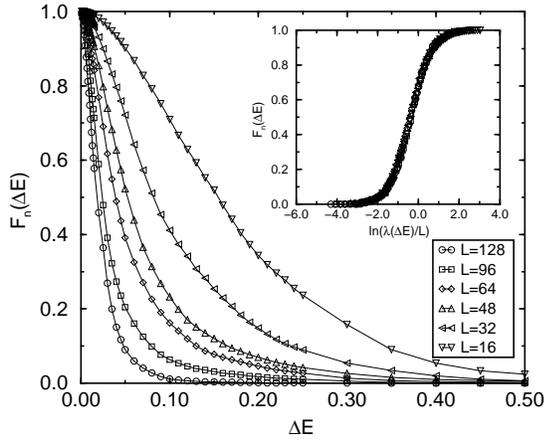}}    
\vspace{-0.5truecm}
\begin{minipage}[t]{8.1cm}    
\caption{ 
The normalized conductance correlation function versus the Fermi energy
difference. The inset shows the scaling plot of the function in 
Eq.~(\ref{nfscale}).
}  
\label{fig3}   
\end{minipage}    
\end{figure}    
The complete behavior of $\lambda(\de)$
is obtained by demanding that all finite size data of $F_{\rm n}$ 
at different $L$ and $\Delta E$ in Fig.~3 collapse onto a single scaling curve
when $F_{\rm n}$ is plotted vs $\lambda/L$. Such a scaling plot is 
shown in the inset of Fig.~3 where the curve represents
the scaling function ${\cal F}$ in Eq.~(\ref{nfscale}). To our
knowledge this is the first demonstration of the scaling behavior
of conductance correlations near a quantum phase transition.
\begin{figure}    
%\hspace{1truecm}    
\vspace{-0.5truecm}
\center    
\centerline{\epsfysize=2.6in    
\epsfbox{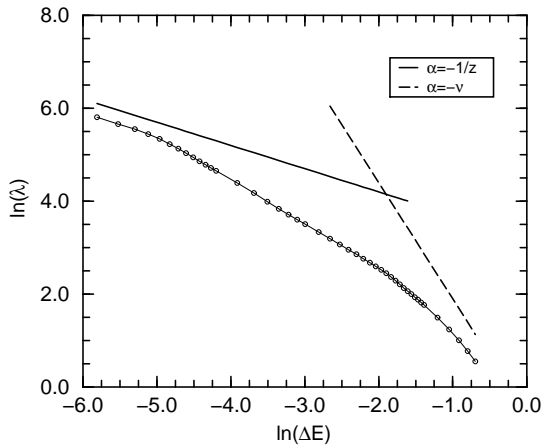}}
\vspace{-0.5truecm}    
\begin{minipage}[t]{8.1cm}    
\caption{ 
The crossover length $\lambda (\de)$ versus $\de$ interpolating
between static and dynamic fluctuation dominated asymptotics.
}  
\label{fig4}   
\end{minipage}    
\end{figure}    
The obtained $\lambda(\de)$ shown
in Fig.~4. Indeed, $\lambda$ interpolates between the
correct asymptotic behaviors dominated by the static and the
dynamical critical fluctuations: (i) For small $\Delta E$,
$\lambda\sim\vert\Delta E\vert^{-1/z}$ and consequently, 
the correlation energy $E_c\sim L^{-z}$. (ii) For large $\Delta E$,
$\lambda\sim\vert\Delta E\vert^{-\nu}$, implying $E_c\sim L^{-1/\nu}$.
(iii) As Fig.~4 shows, the crossover regime between
the asymptotic limits is very broad. Remarkably, over almost the entire
crossover region, $\lambda$ exhibits a well defined power-law,
\eq
\lambda\sim \vert\Delta E\vert^{-1/\zeta},\quad \zeta=0.96\pm0.05.
\label{crossover}
\ee
As a result, the correlation energy $E_c\sim L^{-1}$ over this broad region.
This is consistent with the conclusion drawn from the half-width analysis of
Fig.~3, and provides the microscopic mechanism by which such an unusual 
phenomenon takes place. 

In summary, we have studied the energy correlation 
function of the conductance, a central quantity in mesoscopic physics, 
in the integer QHE. Although we have focused on the QHT in our 
numerical calculations, the physics discussed here is quite general and 
pertains to the critical regimes of quantum phase transitions 
that are driven by the location of the Fermi energy such as
the metal-insulator transitions in disordered electronic systems.

The authors thank David Cobden, Dung-Hai Lee, Nick Read,
Subir Sachdev, and Stuart Trugman 
for useful discussions, and Aspen center for physics for hospitality.
This work is supported in part 
by an award from Research Corporation.
%\end{document}


\begin{references}
\vspace{-1.0truecm}
\bibitem{ucfreview} For reviews, see, {\it e.g.}
{\sl Mesoscopic Phenomena in Solids}, edited by B.~L. Altshuler, P.~A. Lee,
and R.~A. Webb (North-Holland, 1991).
\bibitem{ucf} P.~A. Lee, A.~D. Stone, and H. Fukuyama,
\pprb{\bf 35}, 1039(1987).
\bibitem{cobden}D.~H. Cobden and E. Kogan, \pprb{\bf54}, R17316 (1997).
\bibitem{wjl} Z. Wang, B. Jovanovi\'c, D-H Lee, \pprl {\bf 77}, 
4426 (1996).
\bibitem{chofisher} S. Cho and M.~P.~A. Fisher, \pprb{\bf55}, (1997).
\bibitem{shanhui} S. Xiong, N. Read, and A.~D. Stone, \pprb {\bf56},
3982 (1997).
\bibitem{bhatt} Y. Huo and R. Bhatt, unpublished.
\bibitem{ybkim} H-Y Kee, Y.~B. Kim, E. Abrahams, and R.~N. Bhatt,
cond-mat/9711176.
\bibitem{jw} B. Jovanovi\'c and  Z. Wang, to be published.
\bibitem{huckreview} For reviews, see  B. Huckestein, Rev. Mod. Phys.,
{\bf67}, 357 (1995); and {\sl The Quantum Hall Effect},
eds. R.E. Prange and S. M. Girvin (Springer-Verlag, New York, 1990).
\bibitem{lwinter} D-H Lee and Z. Wang, \pprl {\bf76}, 4014 (1996).
\bibitem{cdnote} J.~T. Chalker and G.~J. Daniell,
\pprl {\bf61}, 593 (1988).
\bibitem{fisherlee} D.~S. Fisher and P.~A. Lee, \pprb{\bf23}, 6851, (1981). 
\bibitem{pichard} J.~L. Pichard and G. Andr\'e, Europhys. Lett. {\bf2},
477 (1986); Y. Imry, Europhys. Lett. {\bf1}, 249 (1986).
\bibitem{pw} V. Plerou and Z. Wang, \pprb,{\bf 58}, 1967 (1998).
\bibitem{cc} J.~T. Chalker and P.~D. Coddington, J. Phys. C {\bf21},
2665 (1988).
\bibitem{lwk}D-H Lee, Z. Wang, and S.~A. Kivelson,
\pprl {\bf70}, 4130 (1993).
D-H Lee, S.~A. Kivelson, Z. Wang,
and S-C Zhang, \pprl {\bf72}, 3918 (1994).
\end{references}
\end{document}